\documentstyle[prb, epsf, psfig,twocolumn,aps]{revtex}

\begin{document}
\draft
\input epsf
\twocolumn[\hsize\textwidth\columnwidth\hsize\csname @twocolumnfalse\endcsname
\title{Thermoelectric Figure of 
Merit of Strongly Correlated Superlattice Semiconductors
}
\author{ Wenjin Mao and Kevin S. Bedell}
\date{\today}

\address{Department of Physics, Boston College, Chestnut Hill, MA
02167, USA}

\maketitle
\widetext 
\begin{abstract}
\leftskip 2cm
\rightskip 2cm

We solved the Anderson Lattice Hamiltonian to get the energy bands of a strongly correlated
semiconductor by using slave boson mean field theory.
The transport properties were calculated in  the relaxation-time approximation, and
the thermoelectric figure of merit was obtained for 
the strongly correlated semiconductor and its superlattice structures. 
We found that at room temperature $ZT$ can reach nearly 2 for the quantum wire lattice 
structure.
We believe that it  is
possible to find high values of thermoelectric figure of merit from  strongly correlated 
semiconductor superlattice systems.

\end{abstract}

 
 \vskip2pc]
\narrowtext
 
For a material to be a good thermoelectric cooler, it must have a high value of 
 the thermoelectric figure of merit $Z$\cite{ZT}, 
which is defined as $ Z=Q^2\sigma/K $, where $ Q $ is the 
 thermopower(Seebeck coefficient), $\sigma$ is the electrical conductivity, and $K$ the thermal 
conductivity.
  The best thermoelectric material now known is $Bi_2Te_3$,
which has $ZT\approx 1$ at room temperature\cite{ZT}. No materials are found at 
lower temperature with $ZT$ larger than 1.
 Recently, much  work has been done on strongly correlated
semiconductors (also termed as Kondo insulators),
 and some authors suggested that this class of
 materials may be good candidates for thermoelectric materials
\cite{Mahanpt,Mahan,Carlos2}. Some experiments have been carried out
to try to
find good thermoelectric materials from strongly correlated semiconductors\cite{Mahanpt,FeSi1},
however, so far there is no satisfactory result.

The strongly correlated semiconductors can be characterized as a
mixed-valence semiconductor with a band gap less than 1000K\cite{Kondo insulators}. 
 The small value of the gap and
its significant temperature dependence have been shown to be consistent with a 
band structure where the hybridization between local $f$ or $d$ electrons 
and the 
conduction band is strongly renormalized by many body effects due to the large on-site $f$ or $d$
 electron Coulomb repulsion\cite{Riseborough,Millis2}. Near the chemical potential the band of 
this class of material is almost flat \cite{Carlos2,Carlos}.
This is 
 a close approximation of the ideal electronic structure 
suggested by Mahan and Sofo \cite{Mahan} for thermoelectric materials, where the transport
distribution function is a delta function centered about 2-3 kT from
the fermi energy\cite{Mahan}.


In this paper we calculate the transport properties of a strongly
correlated semiconductor with an indirect gap of approximately 600K,
which is a good approximation of the condition derived by Mahan and Sofo.
Some authors have argued that a superlattice system might increase
the thermoelectric figure of merit\cite{Hicks}. 
 Here, we also consider the superlattice structure of the strongly 
correlated semiconductor, because the indirect gap for 
them is
proportional to $V^2/W$  
      (where V is the hybridized matrix element and W is the
 band width of the conduction electrons), one may expect a  wider indirect gap
 for a superlattice structure.
 Additionally, the DOS (density
of states) will also be modified to influence 
thermoelectric behavior. We expect that these will  improve the performance of the thermoelectric
 figure of merit.


Our calculation follows the method used by Sanchez-Castro, Bedell, and Cooper
\cite{Carlos2,Carlos}.
 Here the Anderson Lattice Hamiltonian is used, which can be written as:
\begin{eqnarray}
H&=&\sum_{k\sigma}\epsilon_k^0c_{k\sigma}^\dagger c_{k\sigma}+
\sum_{i\sigma}\epsilon_ff_{i\sigma}^\dagger f_{i\sigma} 
  +\frac{1}{2} \sum_{i\sigma\sigma'}Uf_{i\sigma}^\dagger f_{i\sigma}
{f}_{i\sigma'}^{\dagger}{f}_{i\sigma'} \nonumber \\ 
 & &+\frac{1}{\sqrt{N_s} }\ \sum_{ik\sigma} (V e^{-i {\bf k\cdot{R_i}}}
c_{k\sigma}^\dagger f_{i\sigma}+c.c.),
\end{eqnarray}
where $c_{k\sigma}^\dagger$
  is a creation operator for a conduction 
electron with wave vector ${\bf k}$
located 
in the first Brillouin zone and with spin $s=\pm\frac{1}{2}$ ;
$f_{i\sigma}^\dagger
$  is a creation operator for
 a localized $f$ electron centered at ${\bf R_i}$ in the ith unit cell, 
$N_s$ is the number
of unit cells in the crystal. We consider the case $U=\infty$,
 two electrons
 per primitive cell, and degeneracy $N=2$. 
We treat the  $U=\infty$ Anderson Lattice Hamiltonian using a
slave boson formalism. Here
$f^{\dagger}_{i\sigma}$ is represented as the 
bilinear product
 $f_{i\sigma}^\dagger =\widehat{f}_{i\sigma}^\dagger b_i$,
 where $\widehat{f}_{i\sigma}^\dagger $ is a slave fermion creation
 operator representing the $|f_\sigma^{(1)}>_{\bf R_i}$ configuration, 
and $b_i^\dagger $ is a slave boson creation
operator representing the $|f_\sigma^{(0)}>_{\bf R_i}$ configuration. 
In terms of the slave boson operators,
 the $U=\infty$ Anderson lattice Hamiltonian is rewritten as:
\begin{eqnarray}
H'&=&\sum_{k\sigma}\epsilon_k^0c_{k\sigma}^\dagger c_{k\sigma}+
\sum_{i\sigma}\epsilon_f\widehat{f}_{i\sigma}^\dagger 
\widehat{f}_{i\sigma}\nonumber \\
 & &+\frac{1}{\sqrt{N_s} }\ \sum_{ik\sigma} (V e^{-ik\cdot{R_i}}
c_{k\sigma}^\dagger b_i^\dagger \widehat{f}_{i\sigma}+c.c.) \nonumber \\
 & &+\sum_i \lambda_i (\sum_\sigma \widehat{f}_{i\sigma}^\dagger 
\widehat{f}_{i\sigma}
+b_i^\dagger b_i-1)
\end{eqnarray}
where a time independent auxilliary boson field $\lambda_i$ is added, which  couples to the
 system through the last term in the Hamiltonian and enforces the constraint of
no multiple occupancy of a $f$ site. Furthermore, the mean filed 
approximation is applied
to the slave boson 
Hamiltonian by replacing the boson field with expectation
value over the coherent equilibrium states, namely, 
$r=<b_i>$ and $\Lambda=<\lambda_i>$.
Now the resulting Hamiltonian  has
 renormalized
hybridization matrix element $\widetilde{V}=rV$ and renormalized $f$ level 
position $\widetilde{\epsilon}_f=\epsilon_f+\Lambda$.
After changing to the Bloch representation for the $f$ electrons given by 
$\widehat{f}_{k\sigma}^\dagger =\frac{1}{\sqrt{N_s} }
\sum_i e^{-ik\cdot{R_i}} \widehat{f}_{i\sigma}^\dagger ,$
 a canonical transformation is performed to the hybridized band creation
 operators as follows,
 \[  \left[\begin{array}{c}a_{k1\sigma}^\dagger  \\ a_{k2\sigma}^\dagger 
\end{array} \right]=
\left[ \begin{array}{cc}\alpha_{k\sigma}&\beta_{k\sigma}\\
-\beta_{k\sigma}&\alpha_{k\sigma}\end{array}
 \right]
\left[\begin{array}{c}\widehat{f}_{k\sigma}^\dagger 
 \\c_{k\sigma}^\dagger  \end{array} \right]  \]
where $a_{k1\sigma}^\dagger 
= \alpha_k\widehat{f}_{k\sigma}^\dagger  +\beta_k c_{k\sigma}^\dagger $ and
 $a_{k2\sigma}^\dagger 
= -\beta_k\widehat{f}_{k\sigma}^\dagger  +\alpha_k c_{k\sigma}^\dagger $.
\begin{equation}
 \alpha_k^2=\frac{1}{2} (1+\frac{\epsilon_k^0 -\widetilde{\epsilon}_f}{E_k} )
\end{equation}
\begin{equation}
\beta_k^2=\frac{1}{2} (1-\frac{\epsilon_k^0 -\widetilde{\epsilon}_f}{E_k} )
\end{equation}

In terms of the hybridized band operators, the mean field Hamiltonian simplifies to:
\begin{equation}
H^{MF}=\sum_{kn\sigma} \epsilon_{kn} a_{kn\sigma}^\dagger 
 a_{kn\sigma} +N_s \Lambda
(r^2-1),
\end{equation}
where $ n=1,2$ is the  band index and
$\epsilon_{kn}$  is the hybridized band energy given by
\begin{equation}
\epsilon_{kn}=\frac{1}{2} (\epsilon_k^0+\widetilde{\epsilon}_f +
(-1)^{n} E_k),
\end{equation}
and
\begin{equation}
E_k=((\epsilon_k^0 -\widetilde{\epsilon}_f )^2 +4r^2V^2)^{\frac{1}{2} },
\end{equation}

The resulting DOS are given by
\begin{equation}
\rho_{c\sigma}(\omega)=\frac{1}{N_s} \sum_k \delta (\omega-\frac{
\widetilde{V}^2}{\omega-\widetilde{\epsilon}_{f\sigma}}-\epsilon_{k\sigma} )
\end{equation}
\begin{equation}
\rho_{f\sigma}(\omega)=
\frac{\widetilde{V}^2}{(\omega-\widetilde{\epsilon}_{f\sigma})^2}
\rho_{c\sigma}(\omega).
\end{equation}

The equations for determining the 
expectation values of the boson fields are obtained
by minimizing the mean field energy with respect to $ r$ and $\Lambda$ respectively. 
The resulting 
equations to determine $r$, $\Lambda$ and the chemical potential $\mu$ are:
\cite{Lee,Fulde}
\begin{equation}
1-r^2=\frac{1}{N_s} \sum_{k\sigma} (\alpha_{k\sigma}^2 n_F (\epsilon_
{k1\sigma})+\beta_{k\sigma}^2 n_F (\epsilon_{k2\sigma})),
\end{equation}
\begin{equation}
\Lambda=\frac{1}{N_s} \sum_{k\sigma} \frac{V^2}{E_{k\sigma}}
 (n_F(\epsilon_{k1\sigma})-n_F(\epsilon_
{k2\sigma})),
\end{equation}
\begin{equation}
\frac{1}{N_s} \sum_{k\sigma} (n_F (\epsilon_{k1\sigma})+
n_F(\epsilon_{k2\sigma}))=2.
\end{equation}
where $n_F$ is the Fermi-Dirac distribution function.

The relaxation-time approximation\cite{Ashcroft} is used in our calculation,
thus the electrical conductivity $\sigma$, the 
thermopower $Q$, and the thermal conductivity $\kappa$ are given by $\sigma =L_0 $,  
$Q = - \frac{k_B}{|e|} \frac{L_1}{L_0}$, and
$\kappa=\frac{k_B^2 T}{e^2}(L_2-L_1L_0^{-1}L_1),$
where $L_m$ 
is
\begin{equation}
L_m=-\frac{e^2}{V} 
\sum_{kn\sigma} 
\frac{\partial{n_F(\epsilon_{kn\sigma})}}{\partial
{
\epsilon_{kn\sigma}} }
\tau (\epsilon_{kn\sigma}){\bf{v}}_{kn\sigma} {\bf{v}}_{kn\sigma}
(\frac{\epsilon_{kn\sigma}-\mu}{k_B T})^m,
\end{equation}
For the case of a small number of impurities, the electronic  relaxation time is given by
\begin{equation}
\frac{1}{\tau_{kn\sigma}}=\frac{\pi}{\hbar} \frac{N_{imp}}{N_s} |V_0|^2
 (\frac{\partial{\epsilon_{kn\sigma}}}{\partial
{
\epsilon^{0}_{k\sigma}} }
)^2\rho(\epsilon_{kn\sigma}),
\end{equation}
where 
\begin{equation}
\frac{\partial{\epsilon_{kn\sigma}}}{\partial
{
\epsilon^{0}_{k\sigma}} }
=\beta_k^2(\alpha_k^2)
\end{equation}
 for $n=1(2)$ and $N_{imp}$ is the total number of impurities. We have assumed a value of $(N_{imp}/N_s)
{V_0}^2=0.09 eV^2 $ for the impurity potential so that the electrical conductivity
is of the same order as the material FeSi.

Now we calculate the thermoelectric figure of 
merit for bulk material. 
At first, the dispersion relation of the conduction band is assumed to be a tight-binding
band expressed as following:
\begin{equation}
\epsilon_{k\sigma}^0 =\frac{W}{6} (cos (k_xa)+cos(k_ya)+cos(k_za)),
\end{equation}
To perform  our calculation, we consider a cubic lattice with a lattice 
constant $a^3=83 \AA ^3$,
$W=5.0 eV$, 
$V^2=0.4eV^2$, and $\epsilon_f=0.67 eV$ 
below the 
conduction Fermi energy, so that the indirect gap is $610K$ at  $T=0 K$.


In Fig.1 the dimensionless quantity $ZT$ as a function of temperature is shown. 
We find that $ZT$ is no higher than 0.02, and
it is similar to the measurement on FeSi.\cite{Mahanpt}
\begin{figure}
\centerline{\epsfxsize=8cm \epsfysize=6cm \epsfbox{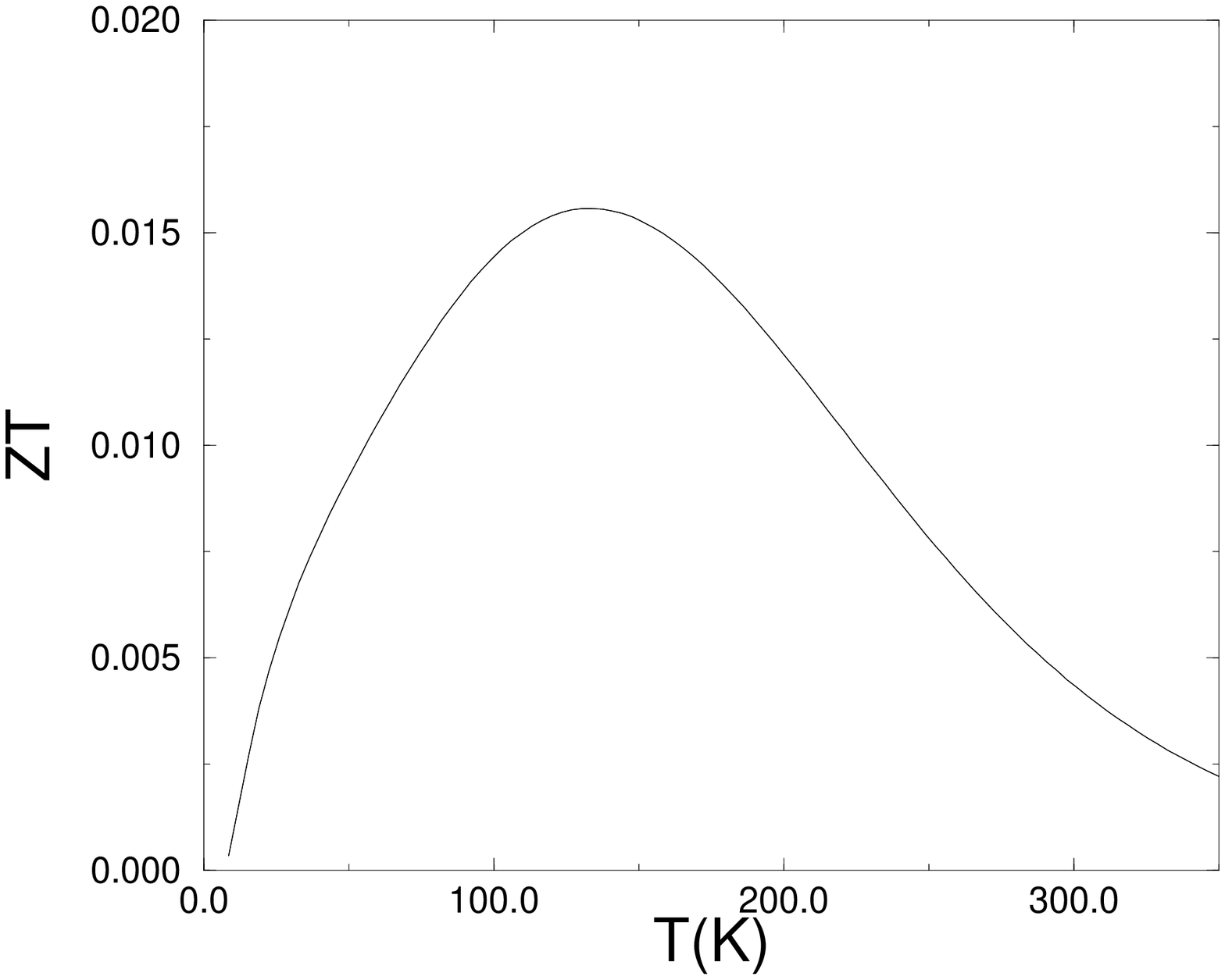}}

\caption{The temperature dependence of the dimensionless quantity 
$ZT$ for bulk material in which the
conduction band is a tight-binding band}
\end{figure}

However, because the electrons and holes are almost evenly distributed when 
a tight-binding model is considered, and the contributions to thermopower
 from holes and electrons are opposite, 
one may expect an increase of the thermopower Q, hence $ZT$, 
if a material has such a conduction band that makes the
 hybridized band more different for holes and electrons respectively.
Thus we 
replace the tight-binding band with a
 nearly parabolic band in our afterwards calculation, and we
will pay more attention on details of this case.

The figure of merit Z has been calculated for (i) a three dimensional (3D) bulk
material, (ii) a 2D multilayered superlattice, and (iii) a 1D quantum wire 
lattice.
 It is assumed that conduction band is  a  nearly parabolic band.
 For a suitably fabricated 
superlattice, the  conduction electrons are confined to move within 2D quantum
 wells or within 1D
quantum wire, they occupy only the lowest
subband of the quantum well or quantum line, and it is approximated that there is  no 
tunneling between layers or lines.
Thus the dispersion relation of conduction electrons in 3D 
is chosen to be
\begin{equation}
\epsilon_k^0=\epsilon_0+\frac{W}{3(\frac{\pi}{a})^2} [f(k_x)+
f(k_y)
+f(k_z)]
 \end{equation}
where
\begin{equation}
f(k)=
\frac{1}{2} (k^2+(\frac{2\pi}{a}-k)^2) 
-[(\frac{(k^2-(\frac{2\pi}{a}-k)^2)}{2})^2+V_0]^\frac{1}{2}
 \end{equation}
and we let $W=6.0eV$, $V_0= 0.1 (\frac{3(\frac{\pi}{a})^2}{W})^2 eV^2$. In the 2D case, the dispersion relation becomes:
\begin{equation}
\epsilon_k^0=\epsilon_0+ \frac{W}{3(\frac{\pi}{a})^2}
[f(k_x)+f(k_y)]
\end{equation}
and in the 1D case , the dispersion relation as:
\begin{equation}
\epsilon_k^0=\epsilon_0+ \frac{W}{3(\frac{\pi}{a})^2}
f(k_x)
 \end{equation}
where $\epsilon_0$ is a constant reference energy. In our calculation, 
we still assume that $\epsilon_f$ is $0.67eV$ below the conduction 
Fermi energy so that the indirect gap at $T=0 K$ is $640 K$ for 3D case.

Fig. 2 shows the transport properties verse temperature for the 3D, 2D and 1D cases, and
Fig. 3 shows results of our calculation of the dimensionless quantity $ZT$ as function of
temperature.
We find that at low temperature, $ZT$ may be very large, however, it has to be noted 
that in our calculation, we have not included lattice thermal conductivity,
 which will dominate the total thermal conductivity at low temperature, thus 
$ZT$ will be dramatically decreased. 
At room temperature, electronic thermal conductivity is 
 larger than or comparable to  lattice thermal conductivity.
What's more, for superlattice system, the phonons can
be scattered by the interfaces and therefore lattice thermal conductivity can be reduced.
Therefore $ZT$ in our calculation
should only be affected a little bit. 
\begin{figure}
\centerline{\epsfxsize=8cm \epsfysize=6cm \epsfbox{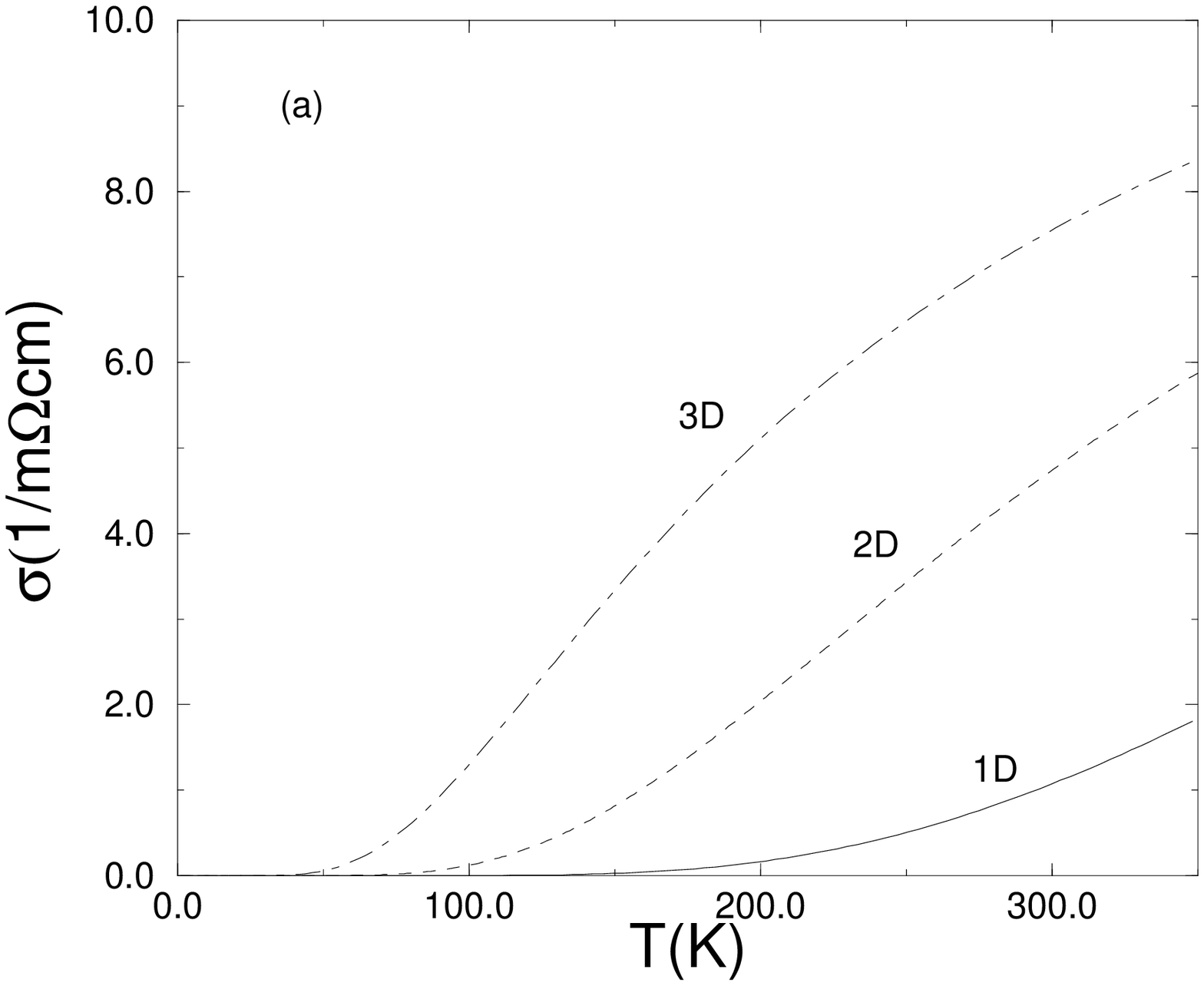}}
\centerline{\epsfxsize=8cm \epsfysize=6cm \epsfbox{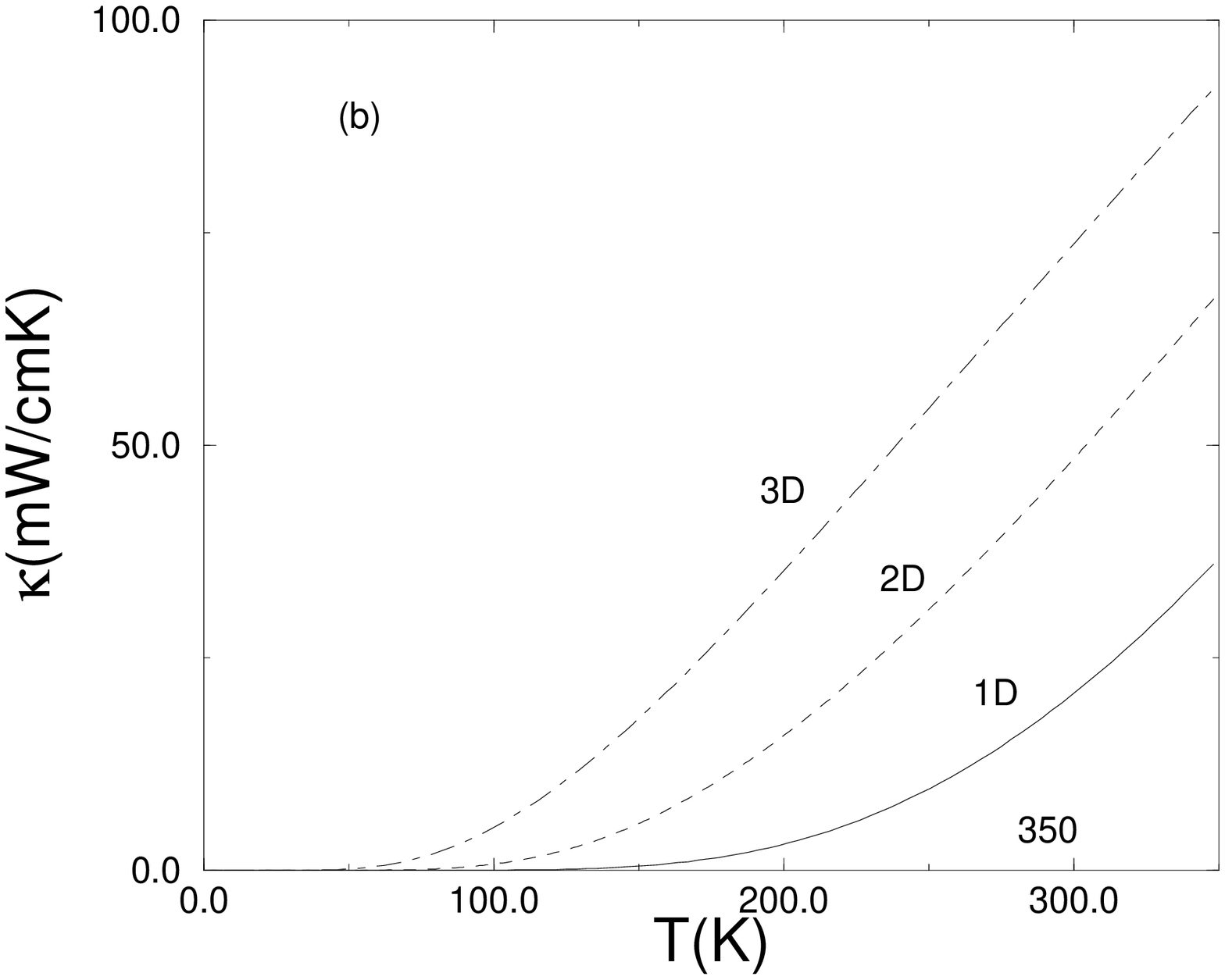}}
\centerline{\epsfxsize=8cm \epsfysize=6cm \epsfbox{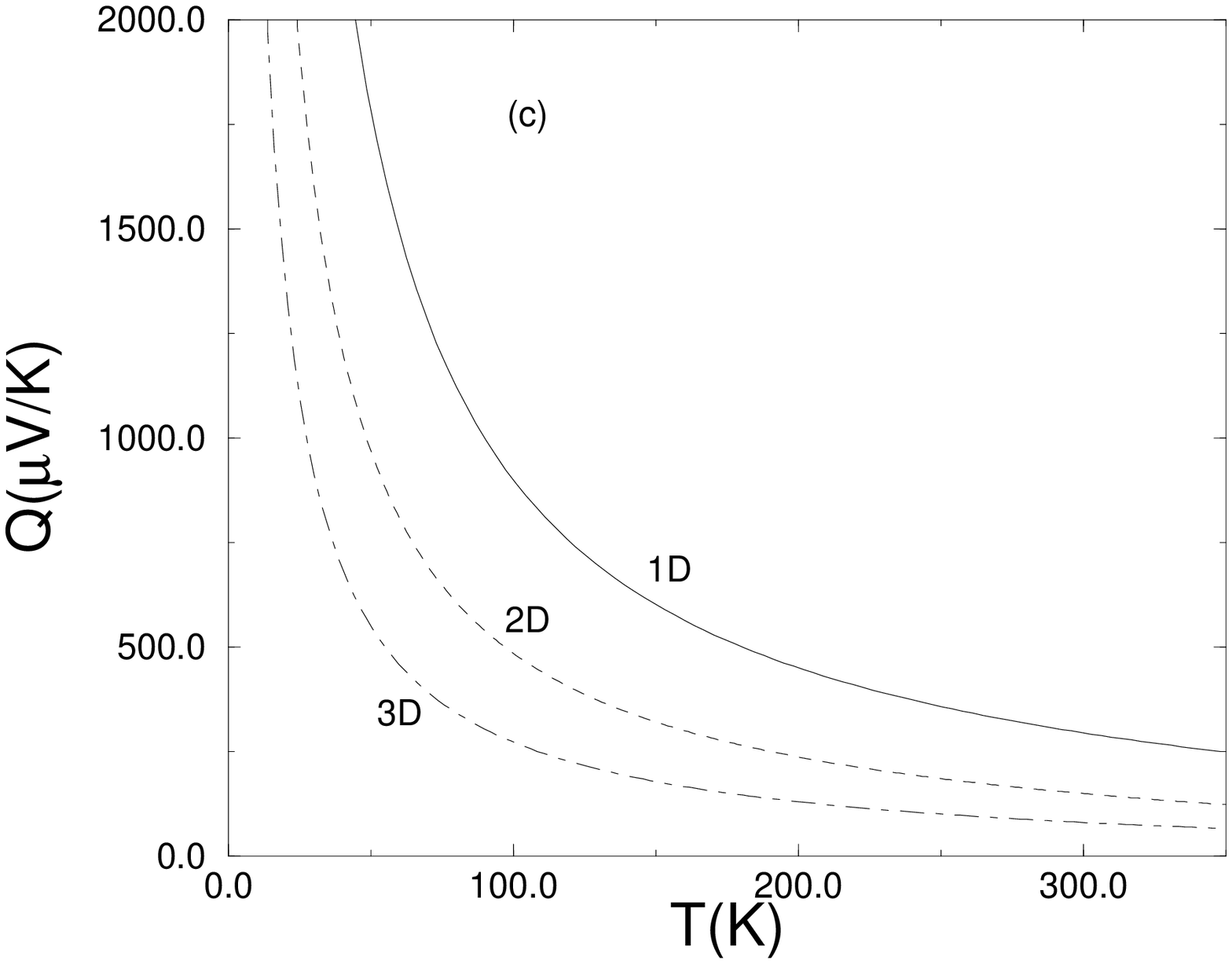}}
\caption{The transport properties (a. electrical conductivity, b. thermal conductivity, c. thermopower) for
3D, 2D and 1D cases in which the
conduction band is nearly parabolic band}
\end{figure}

\begin{figure}
\centerline{\epsfxsize=8cm \epsfysize=6cm \epsfbox{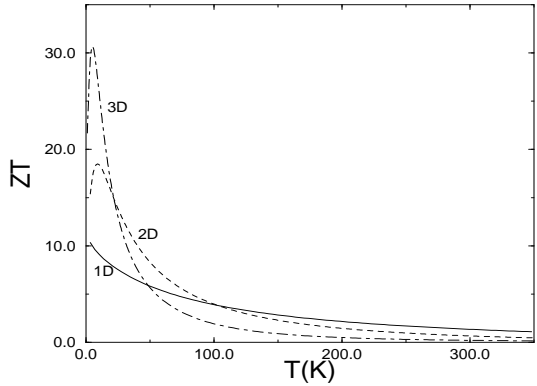}}
\caption{The temperature dependence of the dimensionless quantity $ZT$ for 
3D, 2D and 1D cases in which the
conduction band is nearly parabolic band}
\end{figure}

In Fig. 3 we find that
at room temperature, $ZT=0.2$ for bulk material, 
 $ZT=0.65$ for the case of quantum well, and 
$ZT=1.4$ for quantum wire. We also carried out
 our calculation making the indirect gap of
bulk material  be 1000 K, and found that $ZT$s at room
 temperature for all the three cases
were increased, $ZT$ for 1D reached 2.0, and for 2D nearly 1.5.
In order to compare strongly correlated
 semiconductor with ordinary semiconductors, we made some calculation for ordinary
intrinsic 
semiconductor in the cases of 3D, 2D and 1D. In these calculations, 
we let the indirect gap of the ordinary semiconductor
be the same as the corresponding values of  the strongly
 correlated  semiconductor. At first we adjusted  the effective mass of 
the electron band and the hole band so that in the 3D case $ ZT=0.2$ 
when$ T=300K$, then used these parameter to get $ZT=0.24$ for 2D, and
$ZT=0.4$ for 1D. Comparing with the results for strongly correlated
 semiconductors with $ZT=0.2$, 0.65 and 1.4 for 3D, 2D and 1D cases,
we find there is an advantage in use of strongly correlated semiconductors.

Our calculation indicates that to find good
 thermoelectric material from strongly correlated 
semiconductors, the form of the  conduction band is very important, and it 
 should be a nearly parabolic
band. At low temperature, unless there is some way to 
drop lattice thermal conductivity dramatically, $ZT$ will not be large enough for application. 
At room temperature, for bulk material $ZT$ is still very low, while the quantum wire lattice
structure may have a value of $ZT$ as large as 2.
In addition, it may also be possible to increase $ZT$ by doping 
as in $Bi_2Te_3$, where $Sb$ or $Se$ is added,
and the up to now best p-type or n-type thermoelectric materials are achieved\cite{ZT}.
It is  still interesting to investigate the potential
application of strongly correlated semiconductor  superlattice in the area of thermoelectric cooling, also further improvement by doping as in ordinary semiconductors.

We would like to thank D. A. Broido, J. R. Engelbrecht, Z. Wang, and K. B. Blagoev
for the discussion on this subject. This work was sponsored by the DOE Grant DEFG0297ER45636.

\end{document}